\documentclass[review,onefignum,onetabnum]{siamart220329}
\pdfoutput=1

\usepackage{amsmath}
\usepackage{graphicx}
\usepackage{url}
\usepackage{float}

\usepackage{multicol}
\usepackage{amsmath}
\usepackage{amssymb}
\usepackage{commath}
\usepackage{mathrsfs}
\usepackage{amsfonts}
\usepackage{graphicx}
\usepackage{subcaption}
\usepackage{comment}
\usepackage{bm}
\usepackage{hyperref}

\def\(#1\){{\left(#1\right)}}

\def\N{{\mathbb N}}

\let\ph\varphi

\def\hlink#1{\hbox{\hyperlink{#1}{#1}}}
\graphicspath{{asy/}}

\def\papertitle{Robust Exponential Runge--Kutta Embedded Pairs}
\headers{\papertitle}{T. Zoto and J. C. Bowman}

\title{\papertitle\thanks{Submitted
    to the editors March 25, 2023.\funding{This work was supported by
      the Natural Sciences and Engineering Research Council of Canada.}}}

\author{Thoma Zoto\thanks{RAPSODI, Inria Lille--Nord Europe,
    Villeneuve d'Ascq 59650, France (\email{thomazoto1@gmail.com}).}\and
John C. Bowman\thanks{Department of Mathematical and Statistical Sciences,
  University of Alberta, Edmonton, Alberta T6G 2G1, Canada (\email{bowman@ualberta.ca}).}}

\ifpdf
\hypersetup{
  citebordercolor=0 0 1,
  colorlinks=true,
  pdftitle={\papertitle},
  pdfauthor={Thoma Zoto and John C. Bowman}
}
\fi

\begin{document}

\maketitle

\begin{abstract}
\emph{Exponential integrators} are explicit methods for solving
  ordinary differential equations that treat linear behaviour exactly. 
  The stiff-order conditions for exponential integrators
  derived in a Banach space framework by Hochbruck and Ostermann
  are solved symbolically by expressing the Runge--Kutta weights as
  unknown linear combinations of phi functions.
  Of particular interest are embedded exponential pairs that
  efficiently generate both a high- and low-order estimate,
  allowing for dynamic adjustment of the time step.
  A key requirement is that the pair be \emph{robust:}
  if the nonlinear source function has nonzero total time derivatives,
  the order of the low-order estimate should never exceed its design value.
  Robust exponential Runge--Kutta (3,2) and (4,3) embedded pairs that
  are well-suited to initial value problems with a dominant linearity
  are constructed.
\end{abstract}



\section{Introduction}
Nonlinear initial value problems arise frequently in science and
engineering applications. In contrast with linear first-order ordinary
differential equations, which may be readily solved with the introduction
of an appropriate integrating factor, the solution of nonlinear
initial-value problems typically requires numerical approximation.  

Many dynamical equations contain both linear and nonlinear effects.
In the limit when the linear time scale becomes much shorter than the
nonlinear time scale, it is desirable to solve for the evolution
exactly, for both accuracy and numerical stability. 
For such problems, it may at first seem sufficient to simply
incorporate the linearity into an integrating factor; in the absence
of the nonlinearity, the resulting algorithm would then be exact.
However, such \emph{integrating factor} methods do not solve nonlinear problems
accurately, even on the linear time scale, since the implicit
treatment of the linear term introduces an integrating factor (which
varies on the linear time scale) into the explicitly computed nonlinearity.
Fortunately there exist better methods, known as \emph{exponential}
\emph{integrators}, that are well-suited to such problems: exponential
integrators of order $p$  are exact whenever the nonlinear term is a
polynomial in time of degree less than~$p$.

Consider an initial value problem of the form
\begin{equation}
    \label{eq-LF}
    \frac{dy}{dt} = f(t,y(t)) = F(t,y(t)) - L y, \qquad y(0)=y_0,
\end{equation}
where $y$ is a vector, $F$ is an analytic function, and $L$ is a
constant matrix, with initial condition~$y_0$.
One can compute a numerical approximation of a
future estimate $y_{n+1}$ for $n\ge 0$ using an explicit $s$-stage Runge--Kutta
(RK) method:
\begin{equation}
    \label{eq-RK}
    y_n^{i+1} = y_n^0 +h\sum_{j=0}^ia_{ij}f(t_n+c_jh,y_n^j)
                \quad i=0,\dots,s-1,
\end{equation}
where $y_0^0=y_0$, $y_{n+1}=y_n^s$, $h$ is the {\it time step\/},
$t_n=nh$, the scalar constants~$a_{ij}$ are the Runge--Kutta {\it weights\/},
and $c_j$ are the {\it step fractions\/} for stage $j$. 
For $n \ge 1$, $y_n^0=y_{n-1}^s$ is the approximation of the solution
at time $nh$, also denoted by $y_n$.
The weights can be organized into a
{\it Butcher tableau\/} (Table \ref{t-RK}).
\begin{table}[htbp]
\[
\renewcommand\arraystretch{1.3}
\begin{array}
{c|cccccc}
0\\
c_1
    & a_{00}\\
c_2 
    & a_{10}
    & a_{11}\\
\vdots
    & \vdots
    & \vdots
    & \ddots\\
c_{s-1}
    & a_{(s-2)0}
    & \cdots
    & \cdots
    & a_{(s-2)(s-2)}\\
\hline
1
    & a_{(s-1)0}
    & \cdots
    & \cdots
    & a_{(s-1)(s-1)}\\
\end{array}
\]
    \caption{General Runge--Kutta tableau.}
    \label{t-RK}
\end{table}
For problems where the linear time scale is much shorter than the
nonlinear time scale, explicit Runge--Kutta methods require a very
small time step to maintain stability and accuracy.
This behaviour is often called numerical stiffness and is defined precisely
in \cite{Zoto23schur} and references therein.
Exponential Runge--Kutta (ERK) integrators are explicit methods that
are designed for linearly stiff differential equations.
They have a similar structure to explicit Runge--Kutta methods,
except that the weights $a_{ij}$ are now matrix functions of~$L$:
\begin{equation}
    \label{eq-ERK}
    y_n^{i+1} = e^{-h L}y_n^0 +h\sum_{j=0}^ia_{ij}(-hL) F(t_n+c_jh,y_n^j)
                \quad i=0,\dots,s-1.
\end{equation}
We describe these methods in Section~\ref{exp-integrators}
and introduce in Section~\ref{embedded} computationally
viable embedded exponential pairs that can be used for adaptive time stepping of
practical problems in science and engineering.

We conclude the paper with some numerical examples and applications in
Section~\ref{applications}.

\section{Exponential Integrators}
\label{exp-integrators}
In the scalar case, the global error of the explicit Euler method
grows uncontrollably when $F(t,y)=0$ and $L\tau > 2$. The failure of
explicit methods when $L$ is large is often ascribed in the literature to
\emph{numerical stiffness} \cite{Lambert91}.  First introduced by
Certaine in 1960 \cite{Certaine60},
exponential integrators avoid stiffness arising from the
linear term by treating it exactly.

By defining
$G(t)=F(t,y(t))$, the integrating factor $I(t)=e^{tL}$, and
$Y(t)=I(t)y(t)$, we can transform \eqref{eq-LF} to
\begin{equation}\label{eq-Yt}
    \frac{dY}{dt}=I(t)G(t), \quad Y(0)=y_0.
\end{equation}
Discretizing in the $t$ variable directly leads to
integrating factor methods. However,
the change of variable $I\,dt=L^{-1}\,dI$ transforms~\eqref{eq-Yt} to
\begin{equation}
    \label{eq-YI}
    \frac{dY}{dI}=L^{-1}G(L^{-1}\log(I)).
\end{equation}
Letting $t=t_n+\tau$, we can Taylor expand $G(t_n+\tau)$ about~$t_n$:
\begin{equation}
    G(t_n+\tau) =  \sum_{k=0}^\infty 
                   \frac{\tau^k}{k!}G^{(k)}(t_n),
\end{equation}
so that~\eqref{eq-YI} becomes
\begin{equation}
    \frac{dY}{dI}=L^{-1} \sum_{k=0}^\infty 
    \frac{\(L^{-1}\log(Ie^{-t_nL})\)^k}{k!}G^{(k)}(t_n).
\end{equation}
On integrating from $I(t_n)$ to $I(t_n+h)$, we obtain
\begin{equation}
    Y(t_n+h) = Y(t_n)
    + L^{-1}\sum_{k=0}^\infty G^{(k)}(t_n) \frac{1}{k!}
    \int_{I(t_n)}^{I(t_n+h)}
    \(L^{-1}\log(\tilde I e^{-t_nL})\)^k\,d\tilde I.
\end{equation}
We can change the integration variable from $\tilde I$ to $\bar I=
I(-t_n) \tilde I$ to obtain
\begin{equation}
    Y(t_n+h) = Y(t_n)
    + L^{-1}\sum_{k=0}^\infty G^{(k)}(t_n) \frac{1}{k!}
    \int_1^{I(h)}\(L^{-1}\log \bar I\)^k I(t_n)\,d\bar I
\end{equation}
and then transform back to the original variables, noting $\bar I =
e^{\tau L}$:
\begin{equation}
    e^{(t_n+h)L}y(t_n+h) = e^{t_nL}y(t_n)
    + L^{-1}\sum_{k=0}^\infty G^{(k)}(t_n) \frac{1}{k!}
    \int_0^h\tau^k e^{t_nL} Le^{\tau L}\,d\tau.
\end{equation}
This result simplifies to
\begin{equation}
    y(t_n+h) = e^{-hL}y(t_n)
    + e^{-hL} \sum_{k=0}^\infty G^{(k)}(t_n)
    \int_0^h \frac{\tau^k}{k!} e^{\tau L}\,d\tau.
\end{equation}

We can make this representation of the solution more compact by
defining $\ph_0(x)=e^x$ and
\begin{equation}
    \label{eq-def-ph}
    \ph_k(-hL)=\frac{1}{h^k}\int_0^h \frac{\tau^{k-1}}{(k-1)!}
                e^{-(h-\tau)L}\,d\tau \text{ for } k\in \N.
\end{equation}
Integrating~\eqref{eq-def-ph} by parts leads to an
inductive relation for the $\ph_k$ functions:
\begin{align}
    \ph_k(0) &= \frac{1}{k!},\\
    \ph_0(x) &= e^x,\\
    \ph_{k+1}(x) &= \frac{\ph_k(x)-\frac{1}{k!}}{x} \text{ for }
                    k \geq 0.
\end{align}
The exact solution to~\eqref{eq-LF} then becomes
\begin{equation}
    \label{eq-FL-sol-ph}
    y(t_n+h) = e^{-hL}y(t_n)
            + \sum_{k=0}^\infty 
            h^{k+1} \ph_{k+1} G^{(k)}(t_n),
\end{equation}
which agrees with equation (4.6) of Ref.~\cite{Hochbruck05},
which Hochbruck and Ostermann then use to derive order conditions for
stiff differential equations.
In the literature, \eqref{eq-FL-sol-ph} is typically obtained
with the variation-of-constants method:
\begin{equation}
    \label{eq-FL-sol}
    y(t_n+h) = e^{-hL}y(t_n)
            + e^{-h L}\int_0^h e^{\tau L}F(t_n+\tau,y(t_n+\tau))\,d\tau.
\end{equation}

Since this is the exact solution of~\eqref{eq-LF},
the task of coming up with an exponential Runge--Kutta method
becomes equivalent to approximating the infinite sum in
\eqref{eq-FL-sol-ph} or the integral in~\eqref{eq-FL-sol}.
The simplest approximation of the integral in~\eqref{eq-FL-sol} takes $F$ to be constant: 
\begin{align*}
    y(t_n+h) &= e^{-hL}y(t_n)
            + e^{-hL}\int_0^h e^{\tau L}F(t_n,y(t_n))\,d\tau\\
             &= e^{-hL}y_n + \frac{e^{-hL}-1}{-L}F(t_n,y_n)\\
             &= \ph_0(-hL)y_n + h\ph_1(-hL)F(t_n,y_n).
\end{align*}
The above approximation is called the exponential Euler method;
it reduces to the explicit Euler method in the \emph{classical limit} $L\to 0$.

\begin{figure}[H]
\begin{minipage}{0.49\linewidth}
\centering
\includegraphics[width=\linewidth]{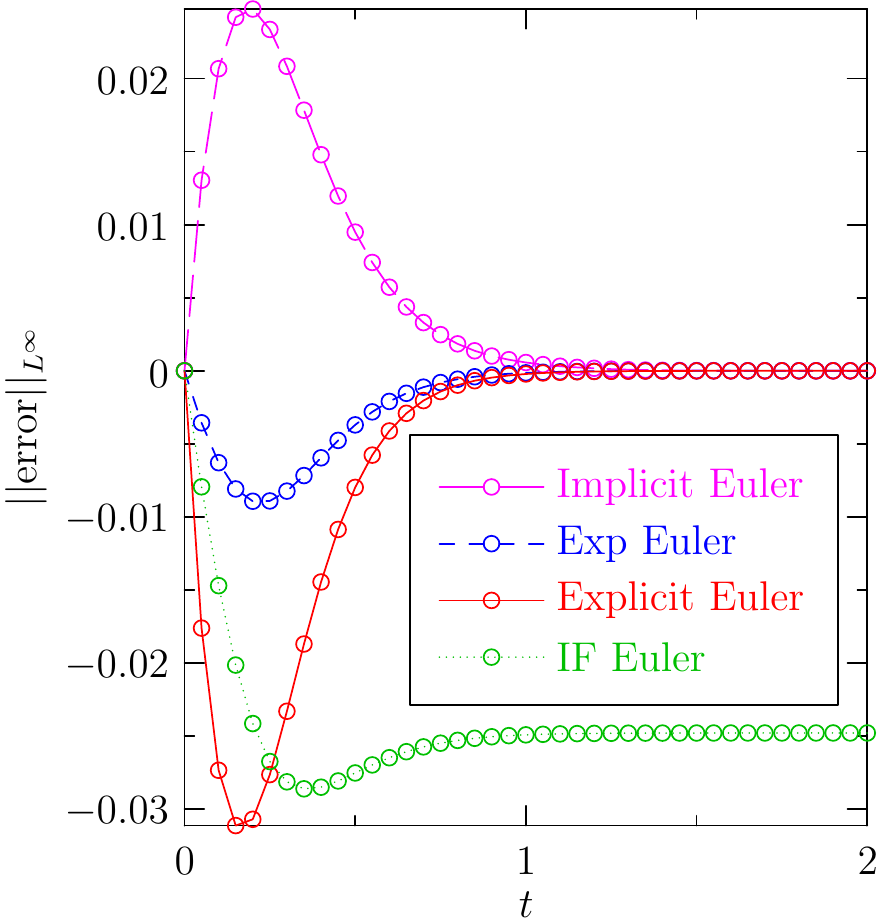}
\caption{Evolution of the error for
   $dy/dt=1/y-Ly$ using the explicit, implicit,
  integrating factor, and exponential Euler methods with $h=0.05$.}
\label{fig-eq-1-over-y-long}
\end{minipage}
\,
\begin{minipage}{0.49\textwidth}
\centering
\includegraphics[width=\linewidth]{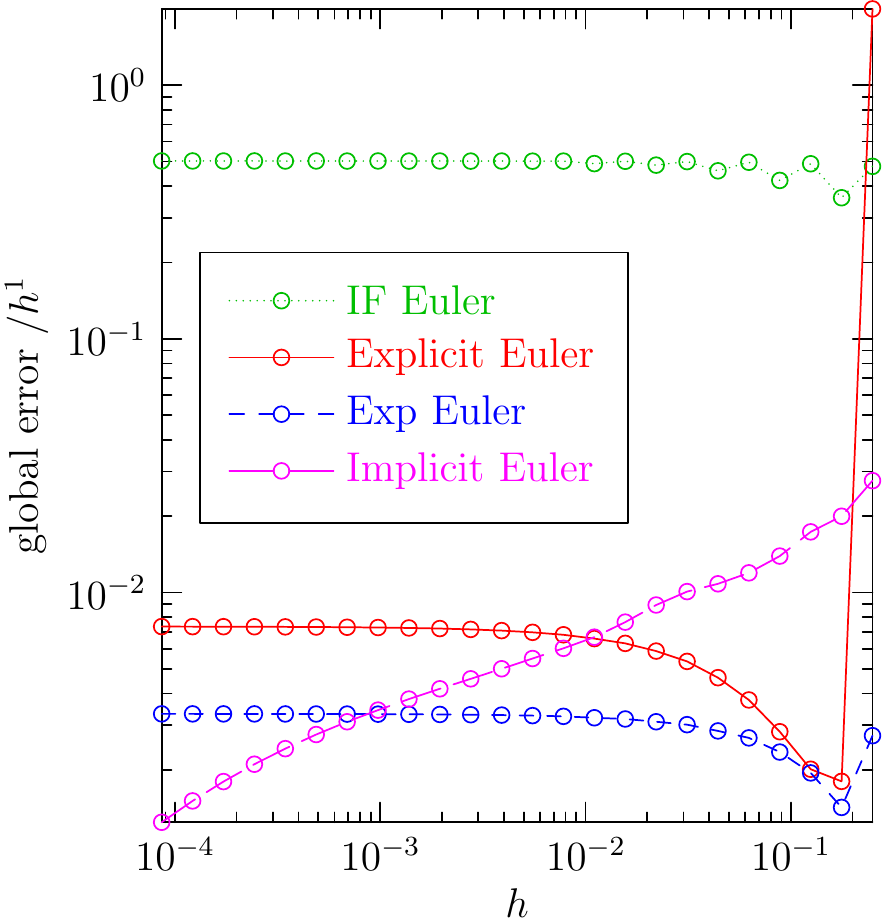}
\caption{Error vs.\ timestep for
   $dy/dt=1/y-Ly$ using the explicit, implicit,
  integrating factor, and exponential Euler methods at $t=1$.}
\label{fig-eq-1-over-y-orders}
\end{minipage}
\end{figure}

The exponential Euler method solves~\eqref{eq-LF}
exactly whenever $F(t,y)$ is constant. In contrast, 
the integrating factor Euler method (IF Euler),
\begin{equation}
    y(t_n+h) = e^{-hL}(y_n + hF(t_n,y_n)),
\end{equation}
which has been widely applied to stiff problems, is exact only when \hbox{$F(t,y)=0$}
and does not preserve fixed points of the original ordinary
differential equation
\cite{Krogstad05,Cox02,Boyd01}, as shown in Figure \ref{fig-eq-1-over-y-long},
where we compare various Euler methods (explicit, implicit, integrator
factor, and exponential), for $F(t,y)=1/y$, with $y_0=1$.
The orders of these methods at $t=1$ are demonstrated in
Figure~\ref{fig-eq-1-over-y-orders}.

Just like in the classical case, there exist high-order
exponential integrator methods. Cox and Matthews \cite{Cox02} attempted to
approximate the integral in~\eqref{eq-FL-sol}
with polynomials of
degree higher than one and subsequently claimed to have derived
a fourth-order scheme that they called ETD4RK.
For the sake of bringing consistency to the names of methods by
different authors, we denote it~\hlink{ERK4CM} and give its
Butcher tableau in Table~\ref{t-ERK4CM}.
\begin{table}[htbp]
  \hypertarget{ERK4CM}\ 
\[
\renewcommand\arraystretch{1.2}
\begin{array}
{c|cccc}
0\\
\frac{1}{2} 
    & \frac{1}{2}\ph_1\(-\frac{hL}{2}\)\\
\frac{3}{4} 
    & 0
    & \frac{1}{2}\ph_1\(-\frac{hL}{2}\)\\
1
    & \frac{1}{2}\ph_1\(-\frac{hL}{2}\)\Bigl(\ph_0\(-\frac{hL}{2}\)
                                                        -1\Bigr)
    & 0
    & \ph_1\(-\frac{hL}{2}\)\\
\hline
1
    & \ph_1-3\ph_2+4\ph_3
    & 2\ph_2-4\ph_3
    & 2\ph_2-4\ph_3
    & 4\ph_3-\ph_2
\end{array}
\]
    \caption{ERK4CM tableau, where $\ph_i=\ph_i(-hL)$.}
    \label{t-ERK4CM}

\end{table}

Krogstad \cite{Krogstad05} takes a different route than Cox and Matthews.
Instead of approximating the integral in~\eqref{eq-FL-sol}, he
truncates the series in~\eqref{eq-FL-sol-ph} in a way that allows
for control of the remainder. He discovered an improved version of
\hlink{ERK4CM} that he claims is fourth order and denotes by ETDRK4-B.
We denote Krogstad's method by~\hlink{ERK4K} and give its Butcher
tableau in Table~\ref{t-ERK4K}.
\begin{table}[htbp]
  \hypertarget{ERK4K}\ 
\[
\renewcommand\arraystretch{1.2}
\begin{array}
{c|cccc}
0\\
\frac{1}{2} 
    & \frac{1}{2}\ph_1\(-\frac{hL}{2}\)\\
\frac{1}{2} 
    & \frac{1}{2}\ph_1\(-\frac{hL}{2}\)-\ph_2\(-\frac{hL}{2}\)
    & \ph_2\(-\frac{hL}{2}\)\\
1
    & \ph_1-2\ph_2
    & 0
    & 2\ph_2\\
\hline
1
    & \ph_1-3\ph_2+4\ph_3
    & 2\ph_2-4\ph_3
    & 2\ph_2-4\ph_3
    & 4\ph_3-\ph_2
\end{array}
\]
    \caption{ERK4K tableau, where $\ph_i = \ph_i(-hL)$.}
    \label{t-ERK4K}
\end{table}

Not long after Krogstad introduced~\hlink{ERK4K},
Hochbruck and Ostermann \cite{Hochbruck05}
published a rigorous framework for deriving exponential integrators.
They argued that simply trying to approximate the integral
in~\eqref{eq-FL-sol}
was not enough to guarantee that the methods
would retain their order for all stiff problems. As
counterexamples, they presented a problem (Fig.~\ref{fig-error-HO}) for
which~\hlink{ERK4CM} exhibits order two (instead of the claimed
order four) and
\hlink{ERK4K} exhibits order three (instead of the claimed order
four). It became apparent that there is a concrete distinction
between the order
conditions for classical and exponential Runge--Kutta methods.
In their Theorem~4.7, Hochbruck and Ostermann derive \emph{stiff-order
  conditions} (Table~\ref{t-stiff}) up to order four that describe
sufficient conditions for an exponential RK method to not suffer a
\emph{stiff-order reduction}.

The operators $J$ and $K$ appearing in the order conditions
in Table~\ref{t-stiff} are discretizations of the bounded operators
\begin{equation}
  J_n = \frac{\partial F}{\partial y}(t_n,y(t_n)), \quad
  K_n = \frac{\partial^2 F}{\partial t \partial y}(t_n,y(t_n)).
\end{equation}
In the case of a single stiff ordinary differential equation, $J$ and $K$, as well as the weights of the
method, are scalars, so that $J$ and $K$ cancel from
the stiff-order equations. If we want a method to retain its
order even when solving arbitrary systems of stiff equations,
we need to be more careful. In this case, the operators $J$ and
$K$ are matrices. Moreover, the weights $a_{ij}$ of the method
are simply linear combinations of $\ph_k$ functions evaluated at
scalar multiples of the matrix $L$.
Therefore, the operators $J$ and $K$ do not
necessarily commute with each other and also do not necessarily
commute with the matrix weights $a_{ij}$. This has to be taken
into consideration when solving the stiff conditions to
construct exponential Runge--Kutta methods.
This delicacy is in fact not restricted to
exponential integrators: the same issue arises when solving order
conditions for classical Runge--Kutta methods.
Butcher gives an example of a classical
RK method that has order five for nonstiff scalar
problems, but only order four when solving nonstiff vector
problems \cite{Butcher09b}.

It is worth noting at this point that although Theorem~4.7 in
Ref.~\cite{Hochbruck05}
is sufficient for an ERK method to retain its order for all stiff
problems, it has not been proven to be necessary. Nevertheless,
Hochbruck and Ostermann show which stiff-order conditions are
violated by~\hlink{ERK4CM},~\hlink{ERK4K}, and some other previously discovered
methods. They proceed to show that there is no method with four
stages that satisfies their stiff-order conditions up to and including
order four. Hence, they introduce a new method that is stiff-order
four, but has five stages. They refer to this method as RK($5.19$),
but for the sake of consistency we denote this method~\hlink{ERK4HO5}
and give its Butcher tableau in Table~\ref{t-ERK4HO5}.
\begin{table}[htbp]
\[
\renewcommand\arraystretch{1.7}
\begin{array}
{|c|c|c|c|c|}
\hline
     \text{No.}
    &\text{order}
    &\text{order condition}\\
\hline
     1
    &1
    &\psi_1(-hL)=0\\
\hline
     2
    &2
    &\psi_2(-hL)=0\\
     3
    &2
    &\psi_{1,i}(-hL)=0\\
\hline
     4
    &3
    &\psi_3(-hL)=0\\
     5
    &3
    &\sum_{i=1}^{s-1}a_{(s-1)i}J\psi_{2,i}(-hL)=0\\
\hline
     6
    &4
    &\psi_4(-hL)=0\\
     7
    &4
    &\sum_{i=1}^{s-1}a_{(s-1)i}J\psi_{3,i}(-hL)=0\\
     8
    &4
    &\sum_{i=1}^{s-1}a_{(s-1)i}J
            \sum_{j=1}^{i-1}a_{ij}(-hL)J\psi_{2,i}(-hL)=0\\
     9
    &4
    &\sum_{i=1}^{s-1}a_{(s-1)i}c_iK\psi_{2,i}(-hL)=0\\
\hline
\end{array}
\]
\begin{align*}
    \psi_{j,i}(-hL) &= \ph_j(-c_ihL)c_i^j
            - \sum_{k=0}^{i-1}a_{ik}(-hL)\frac{c_k^{j-1}}{(j-1)!}\\
    \psi_j(-hL) &= \ph_j(-hL)
        - \sum_{k=0}^{s-1}a_{(s-1)k}(-hL)\frac{c_k^{j-1}}{(j-1)!}
\end{align*}
    \caption{Stiff-order conditions.}
    \label{t-stiff}
\end{table}

\begin{table}[htbp]
  \hypertarget{ERK4HO5}\ 
\[
\renewcommand\arraystretch{1.2}
\begin{array}
{c|ccccc}
0\\
\frac{1}{2} 
    & \frac{1}{2}\ph_1\(-\frac{hL}{2}\)\\
\frac{1}{2} 
    & \frac{1}{2}\ph_1\(-\frac{hL}{2}\) - \ph_2\(-\frac{hL}{2}\)
    & \ph_2\(-\frac{hL}{2}\)\\
1
    & \ph_1-2\ph_2
    & \ph_2
    & \ph_2\\
\frac{1}{2}
    & \frac{1}{2}\ph_1\(-\frac{hL}{2}\)-2a_{31}-a_{33}
    & a_{31}
    & a_{31}
    & \frac{1}{4}\ph_2\(-\frac{hL}{2}\)-a_{31} \\
\hline
1
    & \ph_1-3\ph_2+4\ph3
    & 0
    & 0
    & -\ph_2+4\ph_3
    & 4\ph_2-8\ph_3\\
\end{array}
\]
\begin{align*}
     \ph_i &= \ph_i(-hL),\\
     a_{31} &= \frac{1}{2}\ph_2\(-\frac{hL}{2}\)-\ph_3
     +\frac{1}{4}\ph_2 - \frac{1}{2}\ph_3\(-\frac{hL}{2}\).
\end{align*}
    \caption{ERK4HO5 tableau.}
    \label{t-ERK4HO5}
\end{table}

We have implemented a Mathematica script~\cite{expint} that checks the
stiff order of a method, assuming that we want the method to retain
its stiff order when applied to vector problems.
We use noncommutative algebra in order to prevent Mathematica
from reducing the operators $J$ and $K$.
We verified that the method
\hlink{ERK4K} does not satisfy condition seven and eight in Table
\ref{t-stiff} and hence can exhibit an order reduction to three
in the worst case, confirming the findings of \cite{Hochbruck05}.

Stiff-order conditions up to order
four form a foundation for the derivation of stiff-order conditions
up to order five, tabulated in Table~\ref{t-stiff5}
\cite{Luan14}. Here, the operators $J$ and $K$ are the same as in
Table~\ref{t-stiff}, and $W$ is a discretization of the bounded operator
\begin{equation}
  W_n = \frac{\partial^3 F}{\partial t \partial y^2}
  (t_n,y(t_n)).
\end{equation}
Since the bilinear map $B$ is arbitrary,
in our implementation of these conditions, we 
choose to satisfy condition $17$ in Table~\ref{t-stiff5} by
requiring either $a_{(s-1)i}=0$ or $\psi_{2,i}(-hL)=0$ for
$1\leq i \leq s-1$.

\begin{table}[htbp]
\[
\renewcommand\arraystretch{1.7}
\begin{array}
{|c|c|c|c|c|}
\hline
     \text{No.}
    &\text{order}
    &\text{order condition}\\
\hline
     10
    &5
    &\psi_5(-hL)=0\\
     11
    &5
    &\sum_{i=1}^{s-1}a_{(s-1)i}J\psi_{4,i}(-hL)=0\\
     12
    &5
    &\sum_{i=1}^{s-1}a_{(s-1)i}J
            \sum_{j=1}^{i-1}a_{ij}(-hL)J\psi_{3,j}(-hL)=0\\
     13
    &5
    &\sum_{i=1}^{s-1}a_{(s-1)i}J
            \sum_{j=1}^{i-1}a_{ij}(-hL)J
            \sum_{k=1}^{j-1}a_{jk}(-hL)J\psi_{2,k}(-hL)=0\\
     14
    &5
    &\sum_{i=1}^{s-1}a_{(s-1)i}J
            \sum_{j=1}^{i-1}a_{ij}(-hL)c_jK\psi_{2,j}(-hL)=0\\
     15
    &5
    &\sum_{i=1}^{s-1}a_{(s-1)i}c_iK\psi_{3,i}(-hL)=0\\
     16
    &5
    &\sum_{i=1}^{s-1}a_{(s-1)i}c_iK
            \sum_{j=1}^{i-1}a_{ij}(-hL)J\psi_{2,j}(-hL)=0\\
     17
    &5
    &\sum_{i=1}^{s-1}a_{(s-1)i}B(\psi_{2,i}(-hL),\psi_{2,i}(-hL))=0\\
     18
    &5
    &\sum_{i=1}^{s-1}a_{(s-1)i}c_i^2W\psi_{2,i}(-hL)=0\\
\hline
\end{array}
\]
\begin{align*}
    \psi_{j,i}(-hL) &= \ph_j(-c_ihL)c_i^j
            - \sum_{k=0}^{i-1}a_{ik}(-hL)\frac{c_k^{j-1}}{(j-1)!},\\
    \psi_j(-hL) &= \ph_j(-hL)
        - \sum_{k=0}^{s-1}a_{(s-1)k}(-hL)\frac{c_k^{j-1}}{(j-1)!}.
\end{align*}
    \caption{Stiff-order conditions.}
    \label{t-stiff5}
\end{table}

\section{Embedded exponential pairs}
\label{embedded}

Even when $y$ is a vector and $L$ is a matrix, the weights of
classical Runge--Kutta methods are scalars, while the weights of
exponential integrator methods are linear combinations of
matrix $\ph_k$ functions that depend on the step size $h$ and the matrix
$L$.  This means that we need to re-evaluate
the $\ph_k$ functions whenever $h$ changes.
Since the $\ph_k$ functions involve exponentials
of matrices, variable time stepping is often seen as an expensive
operation when $L$ is a general
matrix \cite{Kassam05}. If $L$ is diagonal, however, the
exponential matrix can obviously be computed efficiently.
Moreover, when $L$ is diagonalizable (e.g.\ if $L$ is normal)
a one-time diagonalization can be used to provide efficient
evaluations of the $\ph_k$ functions for all subsequent values of $h$.
Alternatively, Krylov subspace methods \cite{Tokman10}
can be used to efficiently evaluate the matrix-vector products
that arise in exponential Runge--Kutta integrators.
In designing an integrator, one should keep in mind that choosing many
distinct step fractions $c_i$ requires the evaluation of many
$\ph_k$ functions.

Two embedded ERK methods were introduced by Whalen:
one $(4,3)$ method denoted ETD34 and one $(5,3)$ method denoted
ETD35 \cite{Whalen15}. However, the fourth-order
approximation in ETD34 is the Krogstad method, whose
actual stiff order is $3$. The same holds true for the
fifth-order approximation in ETD35; it is of classical
order five but only of stiff-order three.
Bowman {\it et al.} constructed a stiff $(3,2)$
pair~\hlink{ERKBS32}, given in Table~\ref{t-ERKBS32},
by adding an extra stage to
a special case of a stiff third-order method from
\cite{Hochbruck05}
to obtain a stiff-order two error estimate, such that 
the resulting scheme reduces
to the classical Bogacki--Shampine pair when $L=0$ \cite{Bowman06goy}. 

\begin{table}[htbp]
\[
\renewcommand\arraystretch{1.2}
\begin{array}
{c|cccc}
0\\
\frac{1}{2} 
    & \frac{1}{2}\ph_1\(-\frac{hL}{2}\)\\
\frac{3}{4} 
    & \frac{3}{4}\ph_1\(-\frac{3hL}{4}\) - a_{11}
    & \frac{9}{8}\ph_2\(-\frac{3hL}{4}\)
        +\frac{3}{8}\ph_2\(-\frac{hL}{2}\)\\
\hline
1
    & \ph_1-a_{21}-a_{22}
    & \frac{1}{3}\ph_1
    & \frac{4}{3}\ph_2-\frac{2}{9}\ph_1\\
1
    & \ph_1-\frac{17}{12}\ph_2
    & \frac{1}{2}\ph_2
    & \frac{2}{3}\ph_2
    & \frac{1}{4}\ph_2
\end{array}
\]
    \caption{ERKBS32 tableau, where $\ph_i = \ph_i(-hL)$.}
    \label{t-ERKBS32}
\end{table}

More recently, Ding and Kang
introduced four embedded ERK methods \cite{Ding17}. The first two are built on
the Cox and Matthews method, denoted ERK4(3)3(2), and the Krogstad
method, denoted ERK4(3)3(3), respectively; hence each of them
evidently suffer from order reduction of the higher-order
approximation. The third embedded ERK method,
denoted ERK4(3)4(3) in \cite{Ding17}, is
based on the five-stage method of the stiff-order four
\hlink{ERK4HO5} method.
A stiff-order three method is
appended to~\hlink{ERK4HO5}, resulting in a stiff $(4,3)$ pair.
We denote this pair, given in Table~\ref{t-ERK43DK},
as~\hlink{ERK43DK}.
\begin{table}[htbp]
  \hypertarget{ERK43DK}\ 
\[
\renewcommand\arraystretch{1.2}
\begin{array}
{c|ccccc}
0\\
\frac{1}{2} 
    & \frac{1}{2}\ph_1\(-\frac{hL}{2}\)\\
\frac{1}{2} 
    & \frac{1}{2}\ph_1\(-\frac{hL}{2}\) - \ph_2\(-\frac{hL}{2}\)
    & \ph_2\(-\frac{hL}{2}\)\\
1
    & \ph_1-2\ph_2
    & \ph_2
    & \ph_2\\
\frac{1}{2}
    & \frac{1}{2}\ph_1\(-\frac{hL}{2}\)-2a_{31}-a_{33}
    & a_{31}
    & a_{31}
    & \frac{1}{4}\ph_2\(-\frac{hL}{2}\)-a_{31} \\
\hline
1
    & \ph_1-3\ph_2+4\ph3
    & 0
    & 0
    & -\ph_2+4\ph_3
    & 4\ph_2-8\ph_3\\
1
    & a_{40}
    & \frac{1}{2}a_{44}
    & \frac{1}{2}a_{44}
    & a_{43}
    & 0
\end{array}
\]
\begin{align*}
     \ph_i &= \ph_i(-hL),\\
     a_{31} &= \frac{1}{2}\ph_2\(-\frac{hL}{2}\)-\ph_3
     +\frac{1}{4}\ph_2 - \frac{1}{2}\ph_3\(-\frac{hL}{2}\).
\end{align*}
    \caption{ERK43DK tableau.}
    \label{t-ERK43DK}
\end{table}
The fourth embedded pair is a stiff
$(5,4)$ pair based on a stiff fifth-order method introduced by
Luan and Ostermann \cite{Luan14}. This $(5,4)$ pair is denoted by the original
authors as ERK5(4)5(4), but for the sake of consistency we will
denote it~\hlink{ERK54DK}. 

Exponential integrator methods that converge in the
classical limit $L\to 0$ to a well
studied Runge--Kutta method are attractive. That is why it is
unfortunate that the Krogstad method, which reduces to~\hlink{RK4}
in the
classical limit, does not have stiff-order four
\cite{Hochbruck05}.
This fact implies that there is not a one-to-one
correspondence between RK and ERK methods.
Luan and Ostermann \cite{Luan14} showed that eight stages are
sufficient to achieve stiff-order five. Assuming that their stiff order
conditions are necessary, it appears that an ERK version of
classical $(5,4)$ methods such as \cite{Fehlberg69,Dormand80,Cash90}
is impossible.

\begin{figure}
\begin{minipage}{0.49\linewidth}
    \centering
    \includegraphics[width=\linewidth]{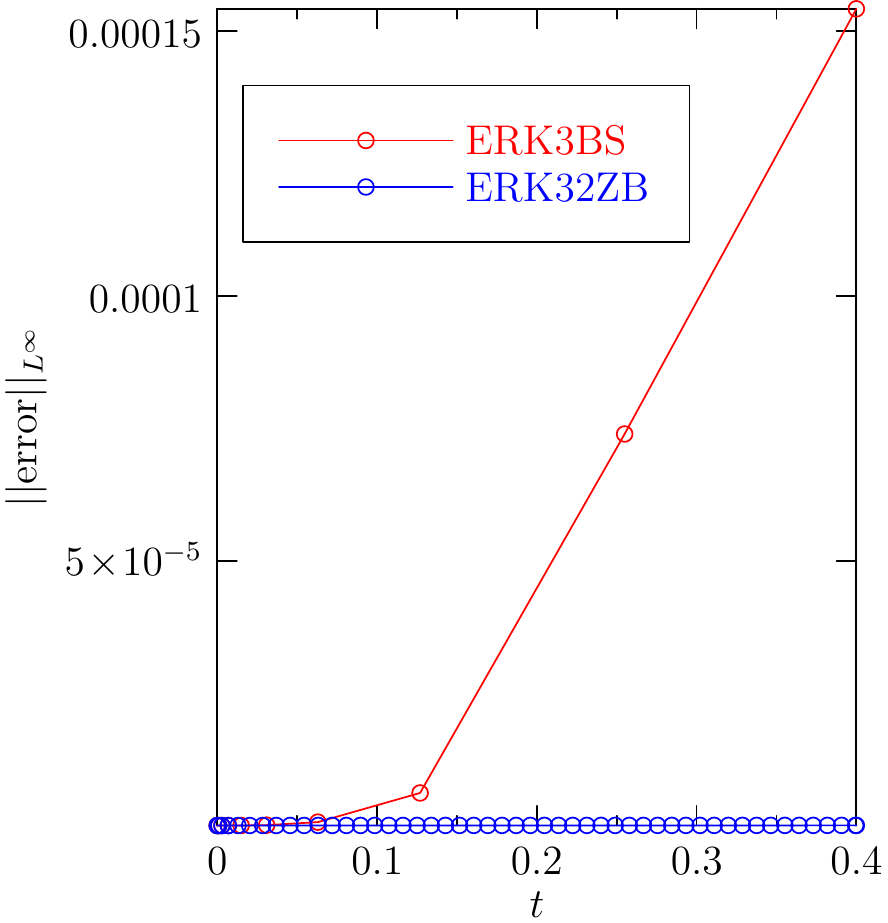}
    \caption{Evolution of the error for~\eqref{eq-HO-6-1}.
      \bigskip\bigskip}
\label{ERK3BSfail}
\end{minipage}
\,
\begin{minipage}{0.49\linewidth}
    \centering
    \includegraphics[width=\linewidth]{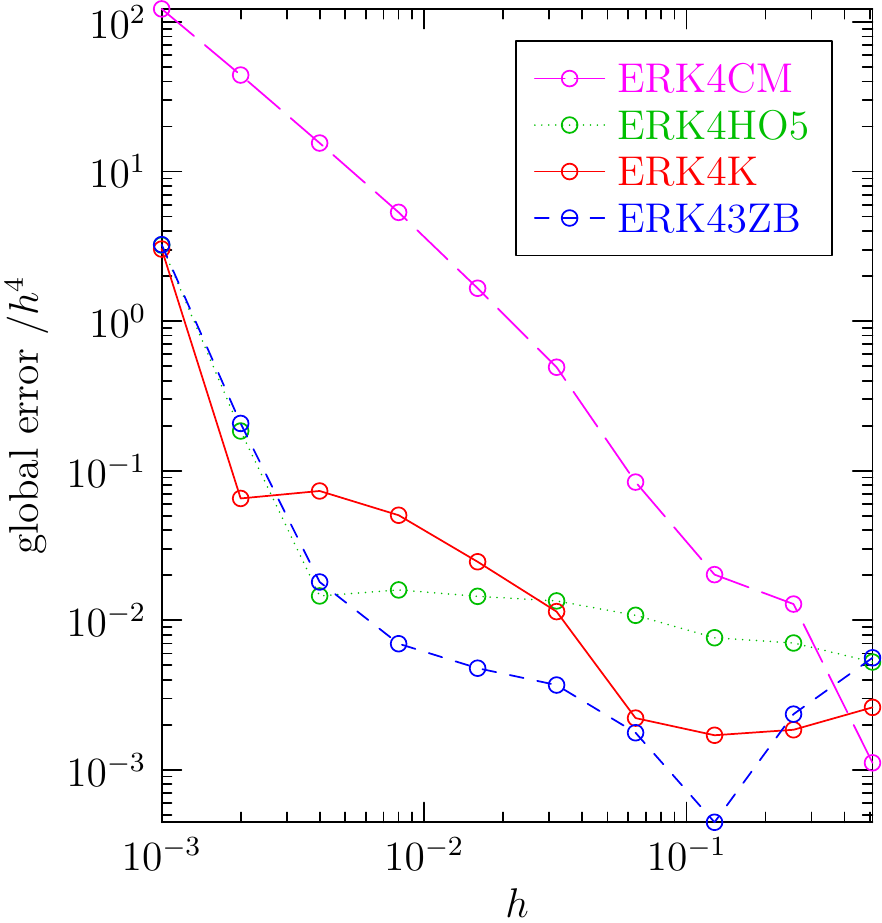}
    \caption{Error vs. timestep in solving~\eqref{eq-HO5} with
      \protect\hlink{ERK4K}, \protect\hlink{ERK4CM},
      \protect\hlink{ERK4HO5}, and \protect\hlink{ERK43ZB} at $t=1$.}
\label{fig-error-HO}
\end{minipage}
\end{figure}

However, the situation might be better than it at first appears.
Usually, to get a classical embedded RK pair
such as the five-stage (4,3) method in Ref~\cite{Balac13},
one adds an extra stage to a classical integrator to obtain an
additional lower-order estimate. Since five is apparently the
minimal number of stages required to achieve stiff-order four, we can
try to use the extra stage to provide an error estimate at no extra cost.
Specifically we seek an embedded ERK method
that gives a stiff third-order approximation in the fourth stage
and a stiff fourth-order
approximation in the fifth stage.
For comparison, the $(4,3)$
stiff pair by Ding and Kang has a total of six stages \cite{Ding17}.
We eliminate the need for a sixth stage by constraining the
second-last stage of the
high-order method to be third order, yielding a $(4,3)$
stiff pair with only five stages, just like in the classical
case. In order to obtain such an embedded method, we need to solve
the system of equations coming from the stiff-order conditions.
We express each weight of the method as
a linear combination of all possible $\ph_k(-c_jhL)$ functions.
By requiring the stiff-order conditions to hold, we are imposing restrictions on
the coefficients of the~$\ph_k$ functions in each weight.
These restrictions form a system of equations
that can be solved symbolically.
The system of restrictions and number of free parameters
becomes relatively large (especially as
we add stages and if we require all of the $c_i$s to be
different).
Provided the symbolic engine can solve the resulting system of restrictions,
it is beneficial if the method has as many free parameters as
possible. We will focus on two 
main areas of optimization. First and foremost, we want the
third-order method to \emph{never} be fourth order for any problem, because
then we would have two fourth-order methods, which would cause the
step-size adjustment algorithm to fail. An example of such a failure
is seen in Figure~\ref{ERK3BSfail} for the problem described by~\eqref{eq-HO-6-1}.

We say that an adaptive pair is \emph{robust} if the order
of the low-order method is never equal to the order $n$ of the
high-order method for any function $G(t)$ with a nonzero
$k$th derivative for some $k\in\{0,\ldots n-1\}$.

To construct a robust pair, we recall
that each ERK method that satisfies the stiff-order conditions
reduces to a Runge--Kutta method that satisfies the classical order
conditions of the same order. However, this does not mean
that the conditions for an ERK method to be robust reduce when $L=0$
to the conditions for a classical method to be robust.
To see this, consider the conditions from Table~\ref{t-stiff} for
a method with four stages to \emph{not} be stiff-order three:
\begin{subequations}
\begin{align}
    \frac{a_{31}c_1^2}{2} + \frac{a_{32}c_2^2}{2} 
    + \frac{a_{33}c_3^2}{2} &\ne \frac{1}{6},\label{degen3a}\\
    c_1^2a_{31}J\ph_2(-c_1hL) + c_2^2a_{32}J\ph_2(-c_2hL)
    + c_3^2a_{33}J\ph_2(-c_3hL) \label{degen3b}\\ \nonumber
    - c_1a_{32}Ja_{11}(-hL)
    - c_1a_{33}Ja_{21}(-hL) - c_2a_{33}Ja_{22}(-hL) &\ne 0,
\end{align}
\end{subequations}
where, consistent with the weak formulation of~
Hochbruck and Ostermann \cite{Hochbruck05},
the weight factors $a_{3j}$ are evaluated at $0$.
The reason that the~\hlink{ERKBS32} method is not robust is
that the equality version of~\eqref{degen3b} in fact holds for all $hL$
and any analytic function $f$ (i.e.\ arbitrary $J$).
Since this degeneracy persists for all values of $h$, an adaptive
time-stepping method will always fail, no matter how much the time
step is adjusted. 
To guarantee that the second-order estimate will never be third-order
for sufficiently small~$h$, it is sufficient to ensure as $hL\to 0$ that
\begin{subequations}
\begin{align}
    \frac{a_{31}c_1^2}{2} + \frac{a_{32}c_2^2}{2} \label{3a-L0}
    + \frac{a_{33}c_3^2}{2} &\ne \frac{1}{6},\\
    \frac{1}{2}c_1^2a_{31} + \frac{1}{2}c_2^2a_{32}
    + \frac{1}{2}c_3^2a_{33} - c_1a_{32}a_{11} - c_1a_{33}a_{21}
    - c_2a_{33}a_{22} &\ne 0.\label{3b-L0}
\end{align}\label{3-L0}
\end{subequations}
But these are not the same as the classical robustness conditions:
\begin{subequations}
\begin{align}
    \frac{a_{31}c_1^2}{2} + \frac{a_{32}c_2^2}{2}\label{3a-cl}
    + \frac{a_{33}c_3^2}{2} &\ne \frac{1}{6},\\
    \frac{1}{6} - c_1a_{32}a_{11} - c_1a_{33}a_{21}
    - c_2a_{33}a_{22} &\ne 0.\label{3b-cl}
\end{align}\label{3-cl}
\end{subequations}

For example, if conditions~\eqref{3-cl} are respected,
it is possible that \eqref{3b-L0} is violated, which when $L=0$ could lead to
the low-order ERK method being third order for some
functions~$f$. To enforce an ERK method to never be third order, one must
respect~\eqref{3-L0}.
In the optimization step we therefore need to
ensure that the third-order estimate does not satisfy any of the
third-order stiff conditions evaluated at $L=0$.

Since we will be advancing the solution using the third-order method,
we would like to minimize its error. To achieve this, we perform a
second optimization: we want the third-order method to be as close
as possible to weakly satisfying the stiff-order four conditions in
\cite{Hochbruck05}.
Moreover, we also need to ensure that the
third-order method, which satisfies the stiff-order three conditions weakly,
is as close as possible to satisfying them strongly.
This is done in a supplementary Mathematica script~\cite{expint}, which generates the
robust exponential $(3,2)$ pair~\hlink{ERK32ZB} displayed in
Table~\ref{t-ERK32ZB}. 

Similarly, we constructed a robust exponential $(4,3)$ pair~\hlink{ERK43ZB},
displayed in Table~\ref{t-ERK43ZB}, by ensuring that the fourth-order
method is as close as possible to weakly satisfying the stiff-order five
conditions in \cite{Luan14} and as close as possible to
satisfying the stiff-order four conditions strongly.
As one can imagine, enforcing these demands is
not as straightforward as when optimizing classical RK methods. The
difficulty lies in the fact that we need to take special care of the
arbitrary operators $J$, $K$, $W$, and the bilinear mapping $B$ that
appear in some of the stiff-order conditions.

\begin{table}
  \hypertarget{ERK32ZB}\ 
\[
\renewcommand\arraystretch{1.2}
\begin{array}
{c|cccccc}
0\\
\frac{1}{2} 
    & \frac{1}{2}\ph_1\(-\frac{hL}{2}\)\\
\frac{3}{4} 
    & \frac{3}{4}\ph_1\(-\frac{3hL}{4}\) - a_{11}
    & \frac{9}{8}\ph_2(-\frac{3hL}{4}) + \frac{3}{8}\ph_2(-\frac{hL}{2})\\
\hline
1
    & \ph_1-a_{21}-a_{22}-a_{23}
    & \frac{3}{4}\ph_2 - \frac{1}{4}\ph_3
    & \frac{5}{6}\ph_2 + \frac{1}{6}\ph_3 \\
1
    & a_{30}
    & a_{31}
    & a_{32}
    & a_{33}
\end{array}
\]
\begin{align*}
    \ph_i  &= \ph_i(-hL)\\
    a_{30} &= \frac{29}{18}\ph_1 + \frac{7}{6}\ph_1\(-\frac{3hL}{4}\)
    +\frac{9}{14}\ph_1\(-\frac{hL}{2}\)
           +\frac{3}{4}\ph_2\\
           &+\frac{2}{7}\ph_2\(-\frac{3hL}{4}\)
           +\frac{1}{12}\ph_2\(-\frac{hL}{2}\)
           -\frac{8083}{420}\ph_3
           +\frac{11}{30}\ph_3\(-\frac{hL}{2}\)\\
    a_{31} &= -\frac{1}{9}\ph_1 - \frac{1}{6}\ph_1\(-\frac{3hL}{4}\)
           -\frac{1}{2}\ph_2\\
           &-\frac{1}{7}\ph_2\(-\frac{3hL}{4}\)
           -\frac{1}{3}\ph_2\(-\frac{hL}{2}\)
           +\frac{1}{6}\ph_3
           +\frac{1}{6}\ph_3\(-\frac{hL}{2}\)\\
    a_{32} &= \frac{2}{3}\ph_1 - \frac{1}{2}\ph_1\(-\frac{3hL}{4}\)
    -\frac{1}{7}\ph_1\(-\frac{hL}{2}\)
           +\frac{1}{3}\ph_2\\
           &-\frac{1}{7}\ph_2\(-\frac{3hL}{4}\)
           -\frac{1}{5}\ph_3\(-\frac{hL}{2}\)\\
    a_{33} &= -\frac{7}{6}\ph_1 - \frac{1}{2}\ph_1\(-\frac{3hL}{4}\)
           -\frac{1}{2}\ph_1\(-\frac{hL}{2}\)
           -\frac{7}{12}\ph_2\\
           &+\frac{1}{4}\ph_2\(-\frac{hL}{2}\)
           +\frac{2671}{140}\ph_3
           -\frac{1}{3}\ph_3\(-\frac{hL}{2}\)
\end{align*}
    \caption{ERK32ZB tableau.}
    \label{t-ERK32ZB}
\end{table}

\begin{table}
\[
\renewcommand\arraystretch{1.2}
\begin{array}
{c|cccccc}
0\\
\frac{1}{6} 
    & \frac{1}{6}\ph_1\(-\frac{hL}{6}\)\\
\frac{1}{2} 
    & \frac{1}{2}\ph_1\(-\frac{hL}{2}\) - a_{11}
    & a_{11}\\
\frac{1}{2}
    & \frac{1}{2}\ph_1\(-\frac{hL}{2}\)-a_{21}-a_{22}
    & a_{21}
    & a_{22}\\
\hline
1
    & \ph_1-a_{31}-a_{32}-a_{33}
    & a_{31}
    & a_{32}
    & a_{33} \\
1
    & \ph_1-\frac{67}{9}\ph_2+\frac{52}{3}\ph_3
    & 8\ph_2-24\ph_3
    & \frac{26}{3}\ph_3-\frac{11}{9}\ph_2
    & a_{43}
    & a_{44}
\end{array}
\]
\begin{align*}
    \ph_i  &= \ph_i(-hL)\\
    a_{11} &= \frac{3}{2}\ph_2\(-\frac{hL}{2}\)
        + \frac{1}{2}\ph_2\(-\frac{hL}{6}\)\\
    a_{21} &= \frac{19}{60}\ph_1 + \frac{1}{2}\ph_1\(-\frac{hL}{2}\)
    +\frac{1}{2}\ph_1\(-\frac{hL}{6}\)\\
           &+2\ph_2\(-\frac{hL}{2}\)
           +\frac{13}{6}\ph_2\(-\frac{hL}{6}\)
    +\frac{3}{5}\ph_3\(-\frac{hL}{2}\)\\
    a_{22} &= -\frac{19}{180}\ph_1
    - \frac{1}{6}\ph_1\(-\frac{hL}{2}\)
    -\frac{1}{6}\ph_1\(-\frac{hL}{6}\)\\
            &-\frac{1}{6}\ph_2\(-\frac{hL}{2}\)
    +\frac{1}{9}\ph_2\(-\frac{hL}{6}\)
    -\frac{1}{5}\ph_3\(-\frac{hL}{2}\)\\
    a_{33} &= \ph_2 + \ph_2\(-\frac{hL}{2}\) - 6\ph_3
            - 3\ph_3\(-\frac{hL}{2}\)\\
    a_{31} &= 3\ph_2 - \frac{9}{2}\ph_2\(-\frac{hL}{2}\)
    - \frac{5}{2}\ph_2\(-\frac{hL}{6}\) + 6a_{33} + a_{21}\\
    a_{32} &= 6\ph_3 + 3\ph_3\(-\frac{hL}{2}\)
    -2a_{33}+a_{22}\\
    a_{43} &= \frac{7}{9}\ph_2 - \frac{10}{3}\ph_3,\qquad
    a_{44} = \frac{4}{3}\ph_3 - \frac{1}{9}\ph_2
\end{align*}
    \caption{ERK43ZB tableau.}
    \label{t-ERK43ZB}
  \hypertarget{ERK43ZB}\ 
\end{table}

As we have already stated,~\hlink{ERK43ZB} is a stiff-order four
method with a second-last stage yielding a stiff-order three estimate
that is guaranteed to never be of higher order. Moreover,
the fourth-order
estimate in~\hlink{ERK43ZB} has minimal error. This has been achieved by
minimizing the $L^2$ norm of a vector $E_5$ whose entries are the coefficients
of every term involving $J$, $K$, $W$, the bilinear
mapping $B$, $\ph_k(-c_jhL)$, and any combination of these
operators. For the method \hlink{ERK43ZB}, $E_5 = 1.43838$.
By comparison,~\hlink{ERK4HO5} has $E_5=6.67545$. This does not mean that the
fourth-order estimate in~\hlink{ERK43ZB}
will do better than~\hlink{ERK4HO5} for every example, but typically
it will be more accurate. As pointed out by
\cite{Dormand80}, there is no need to perform such an
optimization for the lower-order estimate since it is only used for
step-size control.

Accuracy of the higher-order
method is not the only front where the embedded ERK method given by
\cite{Ding17} falls behind. It is not hard to see
that the third-order estimate in
\hlink{ERK43DK} weakly satisfies stiff-order
condition $6$ in Table~\ref{t-stiff}. Hence, there will exist
discretizations for which both estimates will provide a fourth-order
approximation to the solution, as seen in
Figure~\ref{fig-error-too-high}.
This will lead to problems with
step-size adjustment, as seen in
Figure~\ref{fig-error-adaptive}. The same issue arises with the
other stiff pair in \cite{Ding17}: the fourth-order estimate in
\hlink{ERK54DK} weakly satisfies stiff-order condition $10$ in Table
\ref{t-stiff5}. This is also easy to see, noting the similarity
between the last
stage of the fifth-order method and the last stage of the
fourth-order method. There could be
further stiff-order five
conditions that the fourth-order~\hlink{ERK54DK} estimate satisfies,
but this is
enough to show that this embedded method will also cause
step-size adjustment difficulties for some problems.
In contrast, by construction, the~\hlink{ERK43ZB} method does not suffer from
any of these issues.
We demonstrate the new embedded exponential pair~\hlink{ERK43ZB}
for a turbulence shell model in Figure~\ref{ekvkGoy}.

\section{Examples and applications}\label{applications}

We now compare popular exponential integrators with our
proposed integrator~\hlink{ERK43ZB} on the matrix problem in
Example 6.2 of \cite{Hochbruck05}:
\begin{equation}
    \label{eq-HO5}
    \frac{\partial y}{\partial t}(x,t)
    -\frac{\partial^2 y}{\partial x^2}(x,t)
    = \int_0^1 y(\bar x,t)\,d\bar x + \Phi(x,t),
\end{equation}
for $x\in [0,1]$ and $t\in [0,1]$, subject to homogeneous
Dirichlet boundary conditions. The function $\Phi$ is chosen by
substituting in the equation the exact solution which is taken to be
\begin{equation}
    y(x,t) = x(1-x)e^t.
\end{equation}
This problem can be transformed to a system of
ordinary differential equations by performing
a centered spatial discretization of the Laplacian term.
Hochbruck and Ostermann \cite{Hochbruck05} used the discretized version
of \eqref{eq-HO5} with $200$ spatial grid points
to demonstrate the order reduction of the
Krogstad method~\hlink{ERK4K} and the Cox and Matthews method
\hlink{ERK4CM}. We
replicate their results, while adding the error produced
by~\hlink{ERK43ZB}, in Figure~\ref{fig-error-HO}. 
As in \cite{Hochbruck05}, we calculated the matrix
$\ph_k$ functions with the help of Pad\'e approximants. The plots
show the $L^2$-norm of the error at the
time $t=1$. They appear different from the plots in the
original paper only because we divided the error by~$h^4$
(since the methods in consideration are supposed to be fourth order),
allowing us to examine the constant $C$ in front of $h^4$
in the global error. This gives us a better grasp of which
among various methods of the same order is better for the given
problem.

\begin{figure}
\begin{minipage}{0.49\linewidth}
\centering
    \centering
    \includegraphics[width=\linewidth]{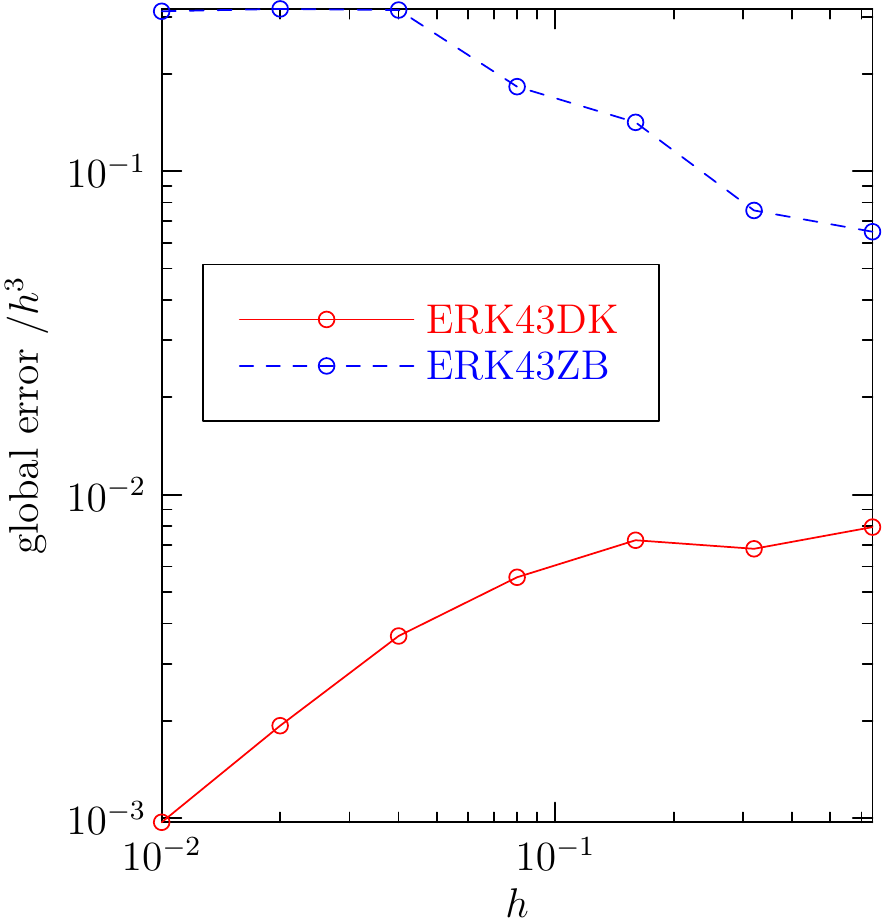}
    \caption{Error vs.\ time step of the third-order estimates
      for~\eqref{eq-HO-6-1} at $t=3$.}
\label{fig-error-too-high}
\end{minipage}
\,
\begin{minipage}{0.49\linewidth}
\centering
    \centering
    \includegraphics[width=\linewidth]{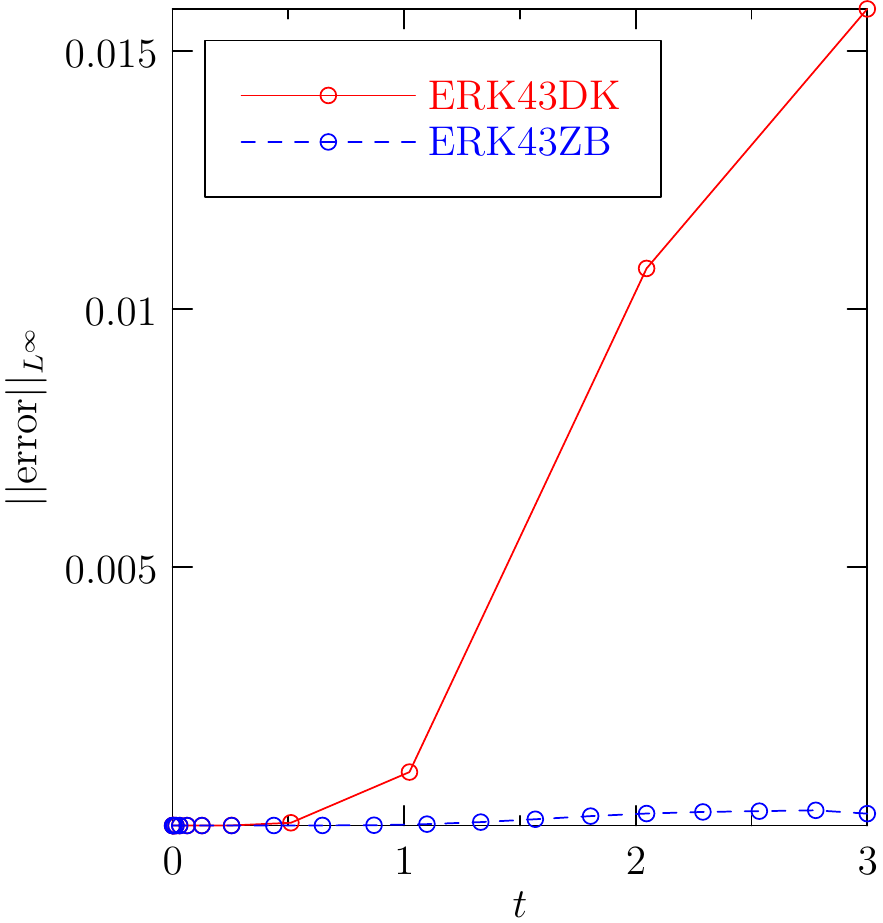}
    \caption{Evolution of the error for~\eqref{eq-HO-6-1}.
      \bigskip\smallskip}
\label{fig-error-adaptive}
\end{minipage}
\end{figure}

We now examine the capability of~\hlink{ERK43ZB} to adjust the step
size as the numerical solution 
is evolved in order for the error to remain within a specified
tolerance. We revisit another example from \cite{Hochbruck05}:
\begin{equation}
    \label{eq-HO-6-1}
    \frac{\partial y}{\partial t}(x,t)
    -\frac{\partial^2 y}{\partial x^2}(x,t)
    = \frac{1}{1+y(x,t)^2} + \Phi(x,t),
\end{equation}
for $x\in [0,1]$. Again, we discretize in space using $200$ grid
points, although this time we continue the time integration from
$t_0=0$ to $t_n=3$ in order to show how~\hlink{ERK43ZB} performs over
a longer time. The exact solution of~\eqref{eq-HO-6-1} is again
taken to be
\begin{equation}
    y(x,t)=x(1-x)e^t
\end{equation}
and the term $\Phi(x,t)$ is calculated by substituting the exact solution
into the equation. Our plan was to compare our stiff $(4,3)$ pair
with the stiff $(4,3)$ pair~\hlink{ERK43DK}, but as we explained at the
end of the previous chapter, the third-order estimate in~\hlink{ERK43DK} is
actually fourth order for some problems, as shown in
Figure~\ref{fig-error-too-high}.
This makes it falsely
conclude that the difference between its fourth-order estimate and
its (supposed) third-order estimate is extremely small, misleading the
time step adjustment algorithm into adopting a time step that is too
large to stay within the specified global error.
In Figure~\ref{fig-error-adaptive} we show the evolution
of the $L^\infty$ error over the time interval $[0,3]$. We note that
\hlink{ERK43ZB} successfully adapts the time step to keep the error small,
whereas~\hlink{ERK43DK} erroneously continues to increase the time step
without bound!

\begin{figure}
\begin{minipage}{0.49\linewidth}
    \centering
    \includegraphics[width=\linewidth]{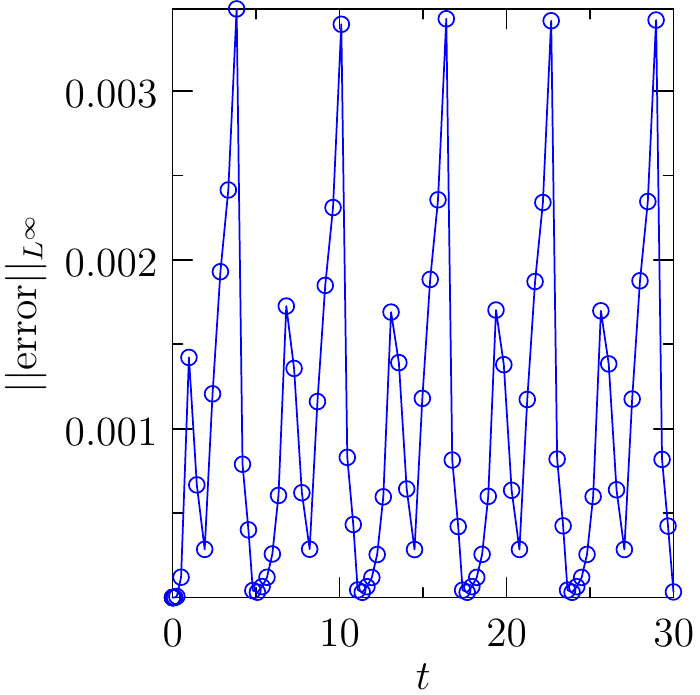}
    \caption{Evolution of the error for~\eqref{eq-HO-6-1} with
      $\Phi$ chosen so that the exact solution is
      $y(x,t)=10(1-x)x(1+\sin t)+2$.
      \bigskip
      }
\label{fig-error-adaptive-sin}
\end{minipage}
\,
\begin{minipage}{0.49\linewidth}
    \centering
    \includegraphics[width=\linewidth]{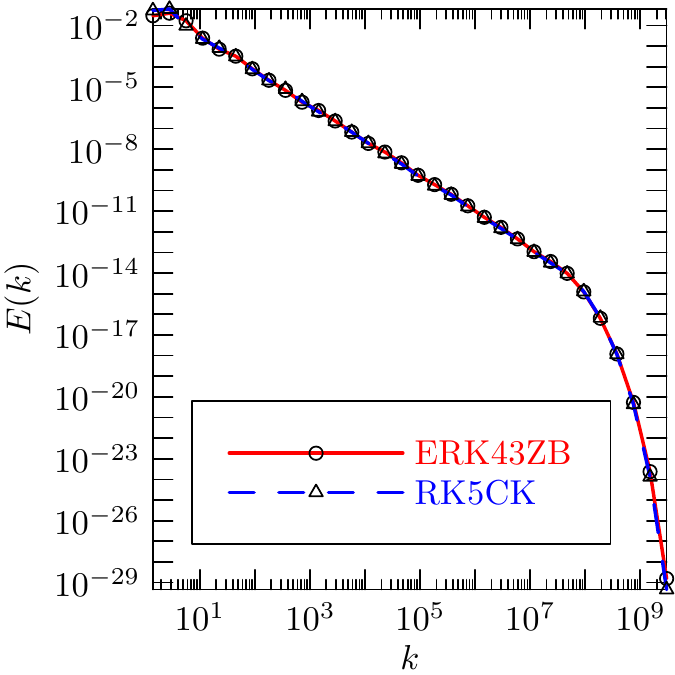}
    \caption{Time-averaged energy spectrum for the GOY shell model
      from $t=5$ to $t=10$ predicted by~\protect\hlink{ERK43ZB} and
      the Cash--Karp (5,4) Runge--Kutta embedded pair RK5CK.
    }
    \label{ekvkGoy}
\end{minipage}
\end{figure}

Perhaps the best way to showcase the ability of~\hlink{ERK43ZB} to
adjust the step size over the course of a long time integration is by
trying to approximate a solution that is periodic in time.
Consider~\ref{eq-HO-6-1} for $x\in [0,1]$ with $\Phi$ chosen so that
the exact solution is $y(x,t)=10(1-x)x(1+\sin t)+2$, discretized
using $200$ grid points and integrated in time from
$t=0$ to $t=30$.
In Figure~\ref{fig-error-adaptive-sin}
we plot the $L^\infty$ norm of the error at each adaptive time step.
The classical Cash--Karp (5,4) Runge--Kutta embedded pair is much less
efficient for this problem: it requires an average timestep roughly
20000 times smaller than that used by~\hlink{ERK43ZB}.

In Figure~\ref{ekvkGoy}, we compare the \hlink{ERK43ZB} embedded pair
to the classical Cash--Karp (5,4) Runge-Kutta pair RK5CK for a
forced-dissipative simulation of the Gledzer--Ohkitani--Yamada
turbulence shell model \cite{Gledzer73,Yamada87,Yamada87b,Yamada88b},
using an adaptive time step, with 32 shells, unit white-noise 
energy injection and Laplacian dissipation coefficient $10^{-11}$.
We benchmarked these algorithms on a single thread of an
5.2GHz Intel i9-12900K processor on an ASUS ROG Strix Z690-F
motherboard with 5GHz DDR5 memory.
The exponential method required 579 seconds, with a mean time step of
$1.39\times 10^{-7}$ seconds, while the classical method required 2098 seconds,
with a mean time step of $3.89\times 10^{-8}$ seconds.
Applying an exponential integrator to this problem, which suffers from
both linear and nonlinear stiffness, led to a speedup of~$3.62$.

\section{Conclusion}

Exponential integrators are well suited to solving linearly stiff ordinary
differential equations. These explicit methods can achieve high
accuracy with relatively large step sizes. In the limit of vanishing linearity, they
reduce to classical discretizations. In this work, we are specifically
interested in robust embedded exponential Runge--Kutta pairs for
adaptive time stepping.
To date, no robust adaptive exponential integrators beyond third order
have been presented in the literature. For example, we have seen that
the exponential version~\hlink{ERKBS32} of the $(3,2)$ embedded Bogacki--Shampine
Runge--Kutta pair \cite{Bowman06goy}, as well as the $(4,3)$
\hlink{ERK43DK} and $(5,4)$ \hlink{ERK54DK} methods in \cite{Ding17},
are not robust.

Symbolic algebra software was used to derive robust exponential
Runge--Kutta (3,2) and (4,3) embedded pairs~\hlink{ERK32ZB} and \hlink{ERK43ZB};
they were optimized to minimize the error of the high-order estimate.
These methods maintain their design orders when applied to
the examples that Hochbruck and Ostermann used to demonstrate stiff
order reduction \cite{Hochbruck05}; they also avoid the step-size
adjustment issues observed with \hlink{ERK43DK} in 
Figure~\ref{fig-error-adaptive}. To illustrate a practical application
of embedded exponential pairs, we showed that the~(4,3) exponential pair
\hlink{ERK43ZB} runs over three times faster than a classical (5,4)
pair when applied to a shell model of three-dimensional
turbulence exhibiting both linear and nonlinear stiffness. 

\section*{Acknowledgment}
Financial support for this work was provided by grants
RES0043585 and RES0046040 from the Natural Sciences and Engineering Research
Council of Canada.

\bibliographystyle{siamplain}
\bibliography{refs}

\begin{thebibliography}{10}

\bibitem{Balac13}
{\sc S.~Balac and F.~Mah{\'e}}, {\em Embedded {R}unge-{K}utta scheme for
  step-size control in the interaction picture method}, Computer Physics
  Communications, 184 (2013), pp.~1211--1219.

\bibitem{Bowman06goy}
{\sc J.~C. Bowman, C.~R. Doering, B.~Eckhardt, J.~Davoudi, M.~Roberts, and
  J.~Schumacher}, {\em Links between dissipation, intermittency, and helicity
  in the {GOY} model revisited}, Physica D, 218 (2006), pp.~1--10.

\bibitem{Boyd01}
{\sc J.~P. Boyd}, {\em Chebyshev and Fourier spectral methods}, Dover, 2001.

\bibitem{Butcher09b}
{\sc J.~C. Butcher}, {\em On fifth and sixth order explicit {R}unge-{K}utta
  methods: order conditions and order barriers}, Canadian Applied Mathematics
  Quarterly, 17 (2009), pp.~433--445.

\bibitem{Cash90}
{\sc J.~R. Cash and A.~H. Karp}, {\em A variable order {R}unge-{K}utta method
  for initial value problems with rapidly varying right-hand sides}, ACM
  Transactions on Mathematical Software (TOMS), 16 (1990), pp.~201--222.

\bibitem{Certaine60}
{\sc J.~Certaine}, {\em The solution of ordinary differential equations with
  large time constants}, Mathematical methods for digital computers, 1 (1960),
  pp.~128--132.

\bibitem{Cox02}
{\sc S.~Cox and P.~Matthews}, {\em Exponential time differencing for stiff
  systems}, J. Comp. Phys., 176 (2002), pp.~430--455.

\bibitem{Ding17}
{\sc X.~Ding and S.~Kang}, {\em Stepsize-adaptive integrators for dissipative
  solitons in cubic-quintic complex ginzburg-landau equations}, arXiv preprint
  arXiv:1703.09622,  (2017).

\bibitem{Dormand80}
{\sc J.~R. Dormand and P.~J. Prince}, {\em A family of embedded {R}unge-{K}utta
  formulae}, Journal of computational and applied mathematics, 6 (1980),
  pp.~19--26.

\bibitem{Fehlberg69}
{\sc E.~Fehlberg}, {\em Klassische {R}unge-{K}utta-formeln f{\"u}nfter und
  siebenter ordnung mit schrittweiten-kontrolle}, Computing, 4 (1969),
  pp.~93--106.

\bibitem{Gledzer73}
{\sc E.~B. Gledzer}, {\em System of hydrodynamic type admitting two quadratic
  integrals of motion}, Sov. Phys. Dokl., 18 (1973), pp.~216--217.

\bibitem{Hochbruck05}
{\sc M.~Hochbruck and A.~Ostermann}, {\em Explicit exponential {R}unge--{K}utta
  methods for semilinear parabolic problems}, SIAM J. Numer. Anal., 43 (2005),
  pp.~1069--1090.

\bibitem{Kassam05}
{\sc A.-K. Kassam and L.~N. Trefethen}, {\em Fourth-order time-stepping for
  stiff pdes}, SIAM J. Sci. Comput., 26 (2005), pp.~1214--1233.

\bibitem{Krogstad05}
{\sc S.~Krogstad}, {\em Generalized integrating factor methods for stiff pdes},
  Journal of Computational Physics, 203 (2005), pp.~72--88.

\bibitem{Lambert91}
{\sc J.~D. Lambert}, {\em Numerical methods for ordinary differential systems:
  the initial value problem}, John Wiley \& Sons, Inc., 1991.

\bibitem{Luan14}
{\sc V.~T. Luan and A.~Ostermann}, {\em Explicit exponential {R}unge-{K}utta
  methods of high order for parabolic problems}, Journal of Computational and
  Applied Mathematics, 256 (2014), pp.~168--179.

\bibitem{Tokman10}
{\sc M.~Tokman and J.~Loffeld}, {\em Efficient design of exponential-{K}rylov
  integrators for large scale computing}, Procedia Computer Science, 1 (2010),
  pp.~229--237.

\bibitem{Whalen15}
{\sc P.~Whalen, M.~Brio, and J.~V. Moloney}, {\em Exponential time-differencing
  with embedded {R}unge-{K}utta adaptive step control}, Journal of
  Computational Physics, 280 (2015), pp.~579--601.

\bibitem{Yamada87}
{\sc M.~Yamada and K.~Ohkitani}, {\em Lyapunov spectrum of a chaotic model of
  three-dimensional turbulence}, J. Phys. Soc. Jap., 56 (1987), pp.~4210--4213.

\bibitem{Yamada88b}
{\sc M.~Yamada and K.~Ohkitani}, {\em Lyapunov spectrum of a model of
  two-dimensional turbulence}, Phys. Rev. Lett., 60 (1988), p.~983.

\bibitem{Yamada87b}
{\sc M.~Yamada and K.~Ohkitani}, {\em Asymptotic formulas for the lyapunov
  spectrum of fully developed shell model turbulence}, Phys. Rev. E, 57 (1998),
  pp.~R6257--R6260, \url{https://doi.org/10.1103/PhysRevE.57.R6257}.

\bibitem{expint}
{\sc T.~Zoto and J.~C. Bowman}, 2023, \url{https://github.com/stiffode/expint}
  (accessed 2023-03-25).

\bibitem{Zoto23schur}
{\sc T.~Zoto and J.~C. Bowman}, {\em Schur decomposition for stiff differential
  equations}, SIAM J. Sci. Comput.,  (2023).
\newblock to be submitted.

\end{thebibliography}
\hypertarget{Bibliography}{}

\end{document}